## Cryptographic Strain-Dependent Light Pattern Generators

*Francesca D'Elia,[1] Francesco Pisani,[2]\* Alessandro Tredicucci,[2,3] Dario Pisignano,[2,3] Andrea Camposeo[3]\**

[1]NEST, Scuola Normale Superiore, Piazza S. Silvestro 12, I-56127 Pisa, Italy

[2]Dipartimento di Fisica, Università di Pisa, Largo B. Pontecorvo 3, I-56127 Pisa, Italy

E-mail: francesco.pisani@df.unipi.it

[3]NEST, Istituto Nanoscienze-CNR and Scuola Normale Superiore, Piazza S. Silvestro 12,

I-56127 Pisa, Italy

E-mail: andrea.camposeo@nano.cnr.it

Abstract.

Refractive freeform components are becoming increasingly relevant for generating controlled patterns of light, because of their capability to spatially-modulate optical signals with high efficiency and low background. However, the use of these devices is still limited by difficulties in manufacturing macroscopic elements with complex, 3-dimensional (3D) surface reliefs. Here, 3D-printed and stretchable magic windows generating light patterns by refraction are introduced. The shape and, consequently, the light texture achieved can be changed through controlled device strain. Cryptographic magic windows are demonstrated through exemplary light patterns, including micro-QR-codes, that are correctly projected and recognized upon strain gating while remaining cryptic for as-produced devices. The light pattern of micro-QR-codes can also be projected by two coupled magic windows, with one of them acting as the decryption key. Such novel, freeform elements with 3D shape and tailored functionalities is relevant for applications in illumination design, smart labels, anti-counterfeiting systems, and cryptographic communication.

## 1. Introduction

Controlling the spatial profile of light beams is critically important in various scientific and technological fields, including high resolution microscopy,[1] endoscopy,[2] lithography and additive manufacturing,[3] optical manipulation of micro-objects,[4] wireless communication[5] and computation.[6] Various methods have been reported to this aim, mostly based on diffractive elements and digital holography, which exploit arrays of micromirrors,[7] liquid crystal-based modulators[8] or metasurfaces.[9]

While such techniques allow high spatial resolution in modulated beams as well as both static and dynamic light patterns to be generated, they typically need highly complex optical elements. This has recently pushed the attention toward refractive freeform optics, that can redistribute the intensity profile from a light source into an arbitrary and pre-determined pattern through simple and robust devices, in which at least one surface has no translational or rotational symmetry with respect to the axis normal to the component main plane.[10] The surfaces of freeform optical elements can be precisely designed in order to produce a desired intensity pattern,[11] defining the involved geometries as sum of spherical or aspherical lenses, or through Q-polynomials description and non-linear partial differential equations.[10,12] The advantages of this method comprise relevant system miniaturization, wider field of view and higher imaging resolution,[2,13,14]. Manifold manufacturing technologies are generally required, involving grinding, polishing, and ultra-precision turning,[15,16] which is highly time-consuming, costly, and poorly versatile, thus preventing freeform optical systems to be realized rapidly and with features variable by external gates. Alternative manufacturing methods are available through 3-dimensional (3D) printing technologies, that can generate objects with unprecedented complex geometries.[17-19] 3D printing comprises a variety of processes to fabricate unconventional architectures by different materials.[20-22] In the field of optics and optoelectronics, additive manufacturing has been already employed to produce aspheric lenses, micro-optics, waveguides, photonic crystals, light-emitting diodes (LEDs), detectors and sensors.[19,23,24] Though 3D printing of macroscopic objects with optical quality and sub-micrometric resolution is still challenging,[25] a number of approaches have been proposed for improving the achievable accuracy, as well as the rate of printing and the size of the printed objects[26-28] Importantly, some applications might exploit light patterns generated from surfaces with lower quality, taking advantage of the design flexibility and customization offered by 3D printing technologies. A relevant example is given by cryptographic labels,[29, 30] where the capability to recognize the generated light patterns by naked eye or by a low-cost scanner, without the need of bulky optics and complex optical systems, is highly desirable.[31, 32] Furthermore, additive manufacturing combined with flexible materials, such as polymeric elastomers, might lead to optical components with high compliance to nonplanar surfaces and large strain.[33] In this respect compression, bending, or stretching, exploited so far for tuning deformable photonic devices,[34-36] could be rethought as effective gating fields to provide 3D optical systems with new functionalities, including controllable properties and cryptographic capability.

Here we introduce 3D printed stretchable *magic windows* (MWs).[37] MWs are transparent refractive components whose surface reliefs are designed by the inverse Laplacian of a target light pattern,[37] and are capable to reshape an incoming light beam into the target image. Our MWs are manufactured by digital light processing (DLP), a fast and cheap 3D printing technology (see Experimental Section for details). The 3D printed MWs, assessed through the achieved spatial distribution of the light intensity, evidence the possibility to design the desired shape of a transmitted optical beam in the whole visible range. Cryptographic systems are demonstrated, in which the information (micro-QR-codes) carried by a light pattern is encrypted in 3D shaped MWs and unveiled by projection of the light pattern through coupled MWs. Moreover, stretchable MWs are made by replica molding of the 3D printed optics, finding that the resulting light pattern can be varied across pre-configured geometries by the application of uniaxial strain to MW elastomers. More specifically, MWs can be designed in order to project indecipherable patterns, whose encoded information (micro-QR-codes) controllably becomes readable upon applying a calibrated uniaxial strain to the devices.

## 2. Results and Discussion

The working principle of MWs is illustrated in **Figure 1**a: it relies on the refraction of light passing through a textured interface between two media with different refractive indices. As schematized in the inset of Figure 1a, the incident light rays are deflected by a certain angle, which depends on the surface texture and the refractive index mismatch. The refracted rays, collected at a defined "*focal*" distance, $f_{MW}$, produce a specific spatial distribution of the light intensity. Given the intensity pattern to be realized, the surface profile of the MW can be calculated by solving the inverse problem, for which various approaches have been reported.[16] Here the method proposed in Ref. [37] is followed: given the desired distance between the plane of projection of the image and the MW ($f_{MW}$), the MW surface profile $h(x,y)$ (where $x,y$ are the in-plane coordinates and $h$ is the thickness of the MW, see Figure 1b) can be calculated by solving the inverse Laplacian of the target light pattern, $I$, (more details about the design and realization of the MWs are reported in the Experimental Section and in Figure S1 of the Supporting Information):[37]

$$\nabla^2 h(R) = \frac{1}{(I(R)-1)f_{MW}(n-1)} \tag{1}$$

where $R=\{x,y\}$, and $n$ is the refractive index ($n=1.5$ in our calculations). From Equation (1), the surface profile $h(x,y)$ is calculated by finite difference approach (see Experimental Section for details) and used for the design of the 3D MW by computer assisted design (CAD).

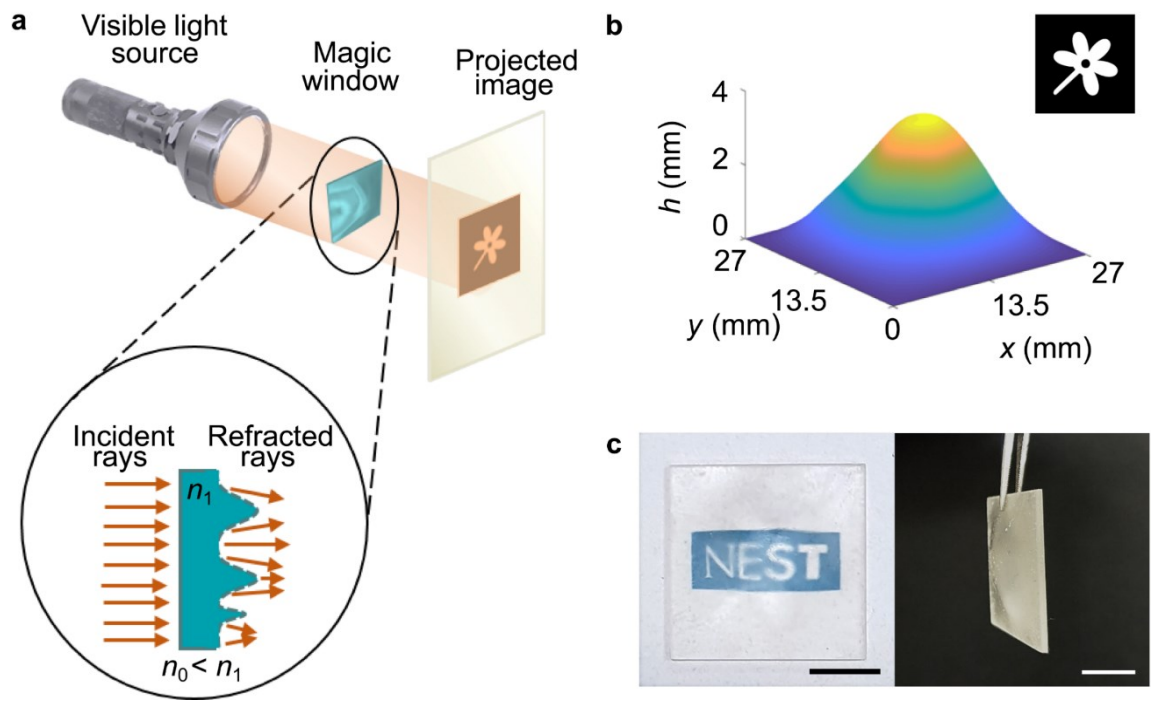

**Figure 1.** 3D printed MWs. **a** Scheme of the MW working principle: light rays from a visible light source are refracted by the MW, whose surface profile is designed in order to generate the target pattern on a screen placed at the focal distance. Inset: scheme of refracted light rays travelling through a textured interface between two media with different refractive indexes ($n_1$ and $n_0$ for the MW and the surrounding medium, respectively). **b** Calculated 3D surface profile of a MW which projects a pattern with a flower (inset). **c** Photographs of a 3D printed MW, showing high transparency on top of the NEST Logo (left), and viewed from its side on black background (right). Scale bars: 1 cm.

Figure 1b shows an example of the calculated surface profile of the MW for the target pattern shown in the inset. The corresponding 3D printed MW is shown in Figure 1c. The MW is almost transparent in the visible range (optical attenuation data are shown in Figure S2 of the Supporting Information) and has no pattern on its surface which is discernible to the naked eye. Importantly, at variance with diffractive architectures and metasurfaces, the overall technology is entirely based on geometrical optics, thus leading to wavelength-independent components made of transparent materials which might exhibit unequalled broadband operation.

Several MWs are designed and printed, with different target patterns: a flower, a Yin Yang symbol, and a square perimeter (**Figure 2**). They are illuminated with various LED sources (Figure S3, and the Experimental Section for details), and the corresponding light patterns projected on a screen are shown in Figure 2, evidencing that the 3D printed MWs can reproduce the desired intensity patterns with high accuracy at several different wavelengths. All images in Figure 2 are obtained at the same focal distance, although this aspect can be in principle affected by the wavelength dispersion of the refractive index of the photo-polymerized MW material (1.48-1.58 for visible light).[38]

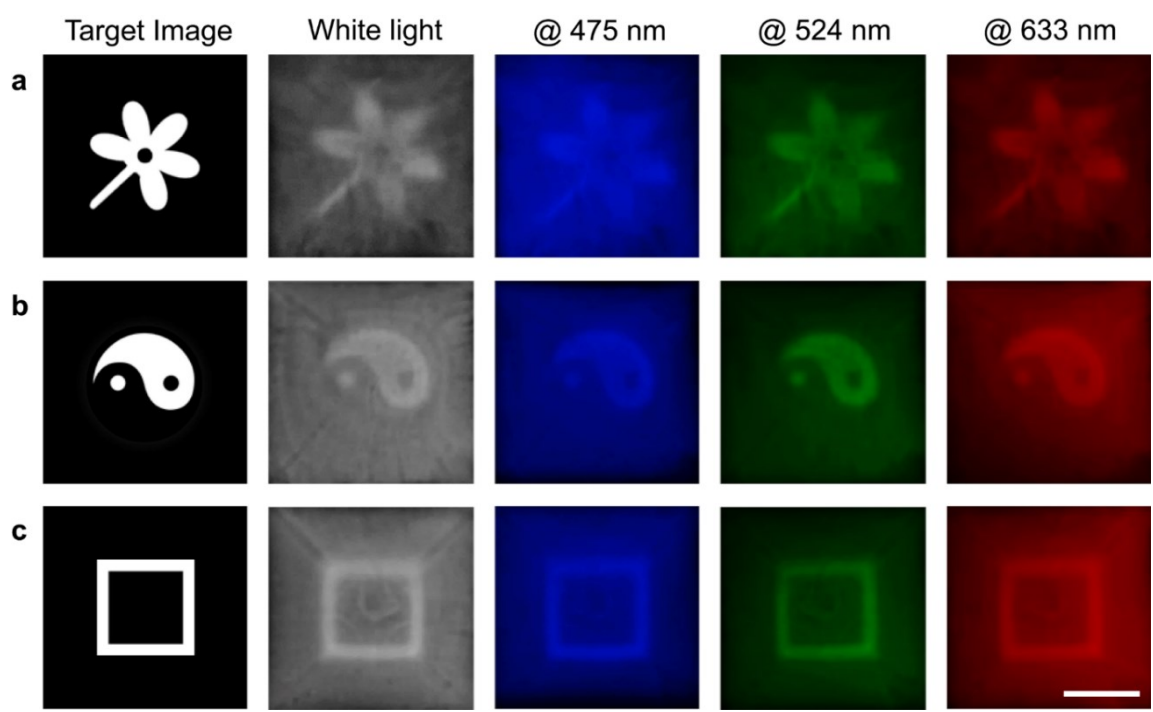

**Figure 2.** Cast images with different visible light sources. **a-c** Target images (left side), and photographs of the images cast by the MWs when illuminated with different LEDs. The images are projected on a screen placed at ~3 mm and captured with a photocamera placed behind the screen. Scale bar: 1 cm.

The variation of the light patterns as a function of the distance, $d$, from the MW is shown in **Figure 3**a-d. Taking the target image as reference (Figure 3a,c), the patterns projected on the screen are better reproduced, in terms of shape and contrast, when the screen is positioned at $d$=3 mm from the MW. The intensity micrographs collected by a CMOS camera (active area 6.66×5.32 mm², Figure 3c,d) is in agreement with the results of the performed ray tracing simulations for various $d$ values (Figure S4).

The properties of the light pattern cast by the MWs are evaluated by three figures of merits: the full width at half maximum (FWHM) of a selected feature (the flower stem in Figure 3), and two contrast parameters, i.e. the contrast-to-noise ratio (CNR) and the Weber contrast (WC), defined as:

$$CNR = \frac{<I_W> - <I_B>}{(\sigma_W + \sigma_B)/2} \qquad (2)$$

$$WC = \frac{<I_W> - <I_B>}{<I_B>} \qquad (3)$$

where $<I_W>$, $<I_B>$, $\sigma_W$ and $\sigma_B$ are the average intensity and standard deviation of light intensity of the white and black regions, respectively (red squares in Figure 3c). WC is a measure of the image contrast normalized to the background luminosity, whereas CNR takes into account also the noise of the image.[39,40] In Figure 3e, the intensity profile of a detail of the projection (red line in Figure 3c) is shown as a function of the distance, $d$. From such data the FWHM of the investigated feature as a function of $d$ is obtained (Figure 3f). The measured size of the flower stem image is found to match the size of the corresponding feature of the target image at $d$=3 mm, in accordance with the value of $f_{MW}$ used for the MW design. Figure 3g shows the $d$-dependence of the contrast parameters, both featuring a maximum at $d$=3 mm, in agreement with the analysis of the feature width. Similar results are obtained for the other studied patterns (Figure S5).

To investigate the minimum feature size that can be projected by the MWs printed by DLP, a set of samples capable of projecting patterns with various geometries (squares and parallel lines) and with features size ranging from 5 mm down to 100 μm are realized (Figure S6 of the Supporting Information). The corresponding projected light patterns upon white light illumination are shown in Figure S6. The analysis of the WC and CNR parameters (Figure S7) performed on arrays of horizontal and vertical parallel lines (Figure S6d), highlights that patterns with feature size down to 300 μm are projected clearly and without appreciable artefacts. The main limitation of the MWs here realized relies on their stepped surface profile (Figure S8 and S9), which impacts more relevantly on features with smaller size which typically have also smaller height (Figure S10). The stepped surface profile of the MWs has the main effect of decreasing the contrast of projected features as shown in Figure S11, where the patterns projected by a smooth and by a stepped surface profile of the MWs are compared. Other parameters of the MWs that affect their height profile are the $f_{MW}$ value, and their lateral size, the thickness of the MW being proportional to the latter and inversely proportional to $f_{MW}$. Such trends allow us to calculate the dependence of the thickness of individual layers for a MW that is sliced in 100 layers as a function of $f_{MW}$ and of the lateral size (Figure S12), and to compare the obtained values with the spatial resolution of various 3D printing technologies. Indeed, current 3D printing methods provide for a wide range of spatial resolutions,[28] and the most suitable technique can be selected depending on the designed MW features. As a general trend, MWs with longer focal distance and smaller lateral size require 3D printing with higher spatial resolution, such as multiphoton stereolitography.[41] Less stringent requirements for spatial resolution are needed for MWs having larger lateral size and shorter focal distances (Figure S12).

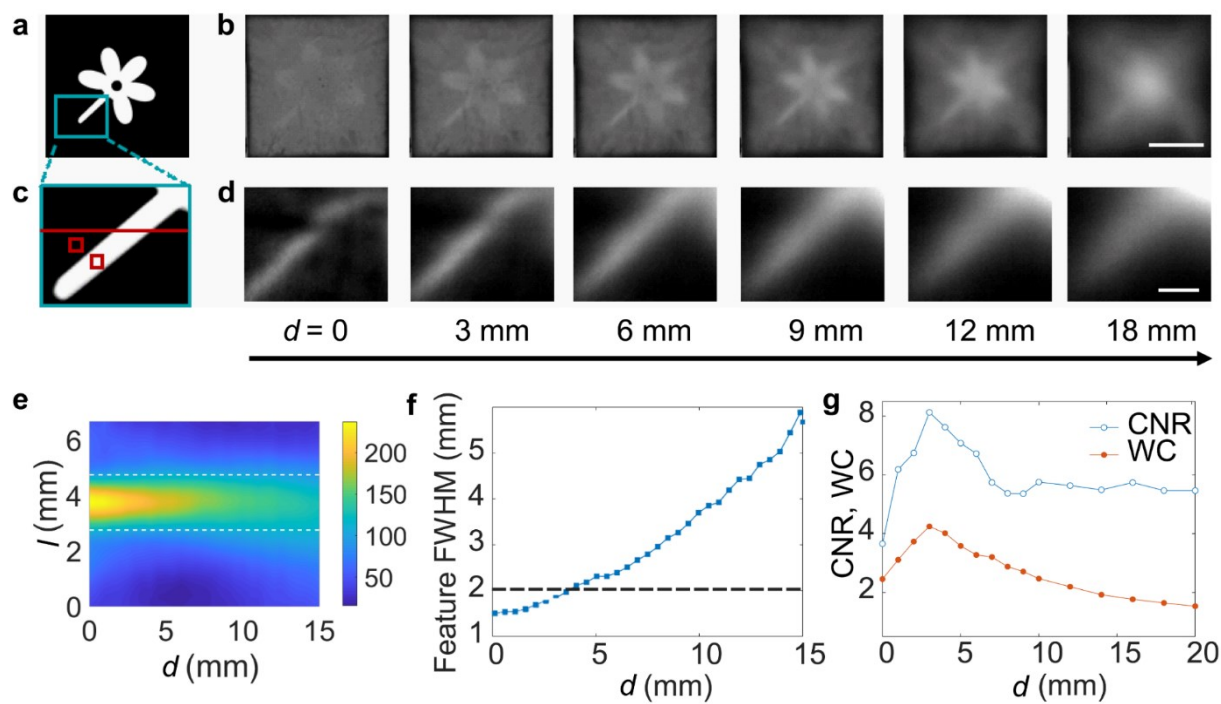

**Figure 3.** Quantitative analysis of MW projection. **a** Target image. **b** Photographs of the projected images at different distances, *d*, of the screen with respect to the MW. *d*=0 corresponds to the MW touching the projection screen. Scale bar: 1 cm. Here blurring and contrast are mostly determined by diffusion and absorption by the semi-transparent screen**. c** Feature chosen for the measurements with the CMOS camera. **d** Maps of the intensity acquired by positioning the CMOS camera at different distances, *d*, from the MW. Scale bar: 2 mm. **e** 2-dimensional map of the light intensity profile *vs*. the distance, *d*, and the coordinate, *l*, along the red line in (**c**). The dashed lines mark the theoretical size of the flower stem. **f** Size of the flower stem (FWHM of the intensity profile) *vs*. *d*. The dashed horizontal line corresponds to the width of the feature of the target image. **g** CNR (empty circle) and WC (full circle) *vs*. *d*. CNR and WC are calculated using the areas highlighted with red boxes in (**c**).

To add a new dimension to freeform optical elements, mechanically deformable MWs are realized made of polydimethylsiloxane (PDMS), by replica molding from the 3D printed MW templates (**Figure 4**a). The PDMS MW has 3D surface reliefs corresponding to the negative of the template, and thus projects an image with black and white areas inverted with respect to the starting 3D printed MW. Therefore, the mold templates are designed and printed to cast patterns with inverted black and white areas with respect to the desired image (Figure 4d,f and Experimental Section for details). The obtained elastomeric MWs project images with lower background compared to the 3D printed ones (Figure 4f *vs*. Figure 4c), because of the reduced bulk light scattering from PDMS that determines lower attenuation in the visible range (Figure S2). Hence, elastomeric MWs generally outperform 3D printed templates in terms of the obtained image quality and contrast (Figure S6 and S7). A variation of the patterns generated from the PDMS MW similar to the 3D printed one is found upon increasing the distance, $d$ (Figure S13).

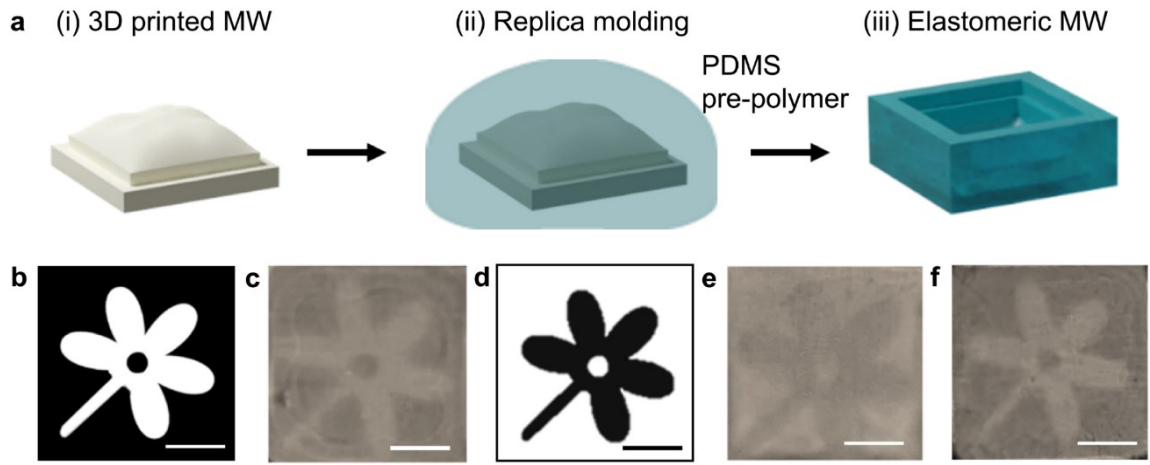

**Figure 4.** Elastomeric MWs. **a** Schematic view of the replica molding process of the elastomeric MWs. A template is 3D printed (i) and covered with PDMS pre-polymer (ii), which is then cured. The resulting elastomeric MW (iii) is peeled off from the substrate. **b** Target image and **c** corresponding image projected by the 3D printed MWs. **d** Target image with inverted black and white areas with respect to the one shown in (**b**). **e-f** Corresponding images projected by the 3D printed MW (**e**) and by the PDMS replica (**f**), realized starting from (**e**). Scale bars: 1 cm.

MWs have the interesting property to unveil a structured pattern on a screen positioned at a well-defined distance ($f_{MW}$) when illuminated. We exploit this property by realizing cryptographic MWs designed to project the complex pattern of a micro-QR-code encoded with the letters "ABVZ" (**Figure 5**a). The micro-QR-codes can be encrypted in the topography of MWs through the calculation of the inverse Laplacian, while the feature of the topography of both printed and elastomeric MWs (almost invisible to naked eyes) cannot reveal the code content (Figure 5b). The reading of the QR-codes is then carried out by illuminating the MWs and projecting the crypted pattern on a screen at $f_{MW}$ distance (Figure 5c).

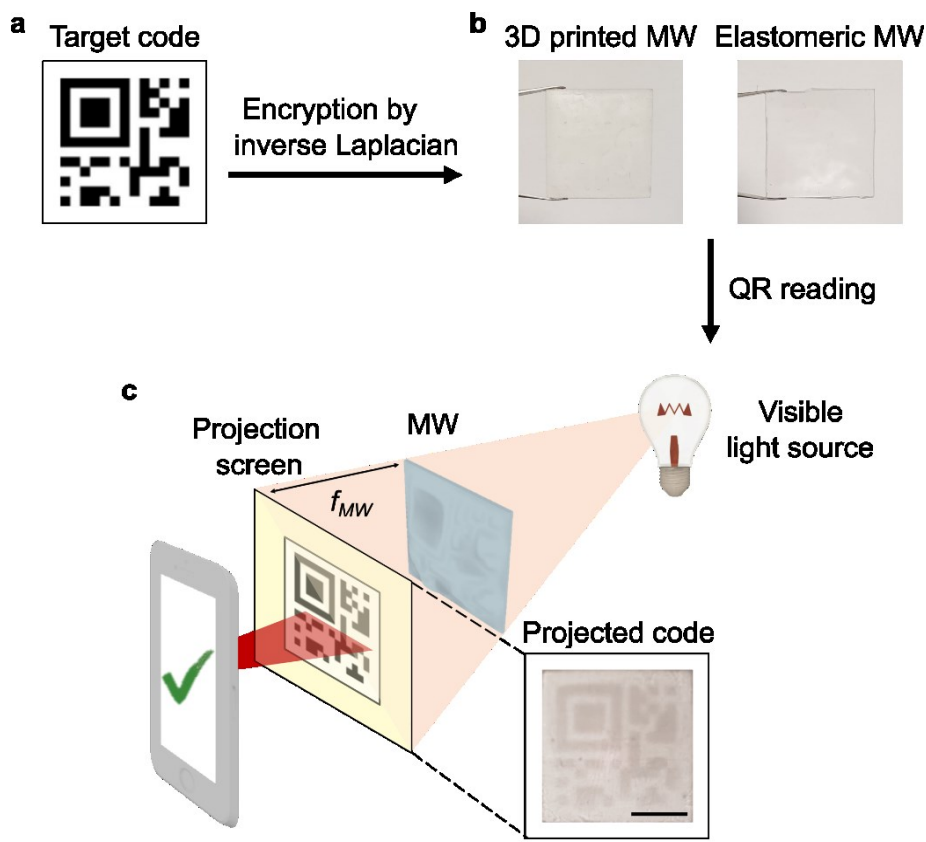

**Figure 5**. **a** Target image of a micro-QR-code encoded with the letters "ABVZ". **b** Photographs of the MWs realized by 3D printing (left image) and replica molding (right image) methods. MW lateral size = 27×27 mm$^2$. **c** Scheme of the QR reading mechanism based on the illumination of the MW by a broadband light source and the projection of the crypted pattern on a screen. The projected pattern can be read by a QR-scanner. The inset shows an example of the pattern projected by a MW realized starting from the image shown in **a**. Scale bar: 1 cm.

The above encryption mechanism can be further strengthened by utilizing two coupled MWs, whose encrypted light pattern can be unveiled only if the two optical components are placed in series, while remaining undecipherable if only one of them is used (**Figure 6**). In such approach, one of the two MWs acts as unique key for decryption.

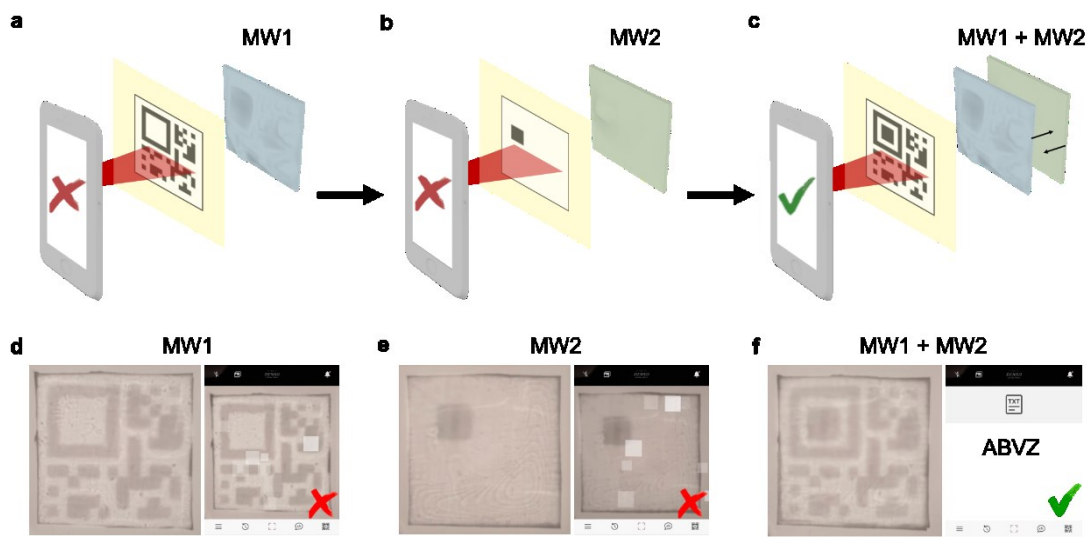

**Figure 6**. Cryptographic MW with decryption key. **a,c** Schematic illustration of the operation of two coupled cryptographic MWs (MW1 and MW2). The projected pattern is unreadable when a single MW is used (**a**, **b**), while becoming readable when they are in series (**c**). Photographs (left side of each panel) of the patterns projected by the cryptographic elastomeric MW1 (**d**), MW2 (**e**) and by the combination of the two (**f**). On the right side of **d**-**f** the screenshot of the QR scanner is shown. The pattern of gray squares in (**e**),(**f**) is due to the graphic interface of the QR scanner.

Finally, the surface morphology of elastomeric MWs can be varied by mechanical elongation, thus opening the possibility to modify the resulting light pattern generated by controlling the device stretching. This property can be exploited as an additional degree of encryption of the information carried by the MW. To this aim, elastomeric MWs are designed starting from a pattern of the micro-QR code that is intentionally encrypted in a way that it cannot be read by a QR scanner unless a pre-defined uniaxial strain, $\varepsilon = \Delta L / L_0$, (where $L_0$ is the lateral size of the unstretched MW) is applied (**Figure 7**a,b). We realize two MWs textured with different surface profiles, namely designed to work upon calibrated elongation of 15-20% (Figure 5c) and of 20-30% (Figure 7d), respectively. Slight reductions of the set elongation points might be related to the algorithms exploited for error correction by QR scanners, and to possible damage or deformation compensation of the 2-dimensional code pattern.[42,43] An exemplary behavior of the MW under stretching is also shown in the Supporting Movie.

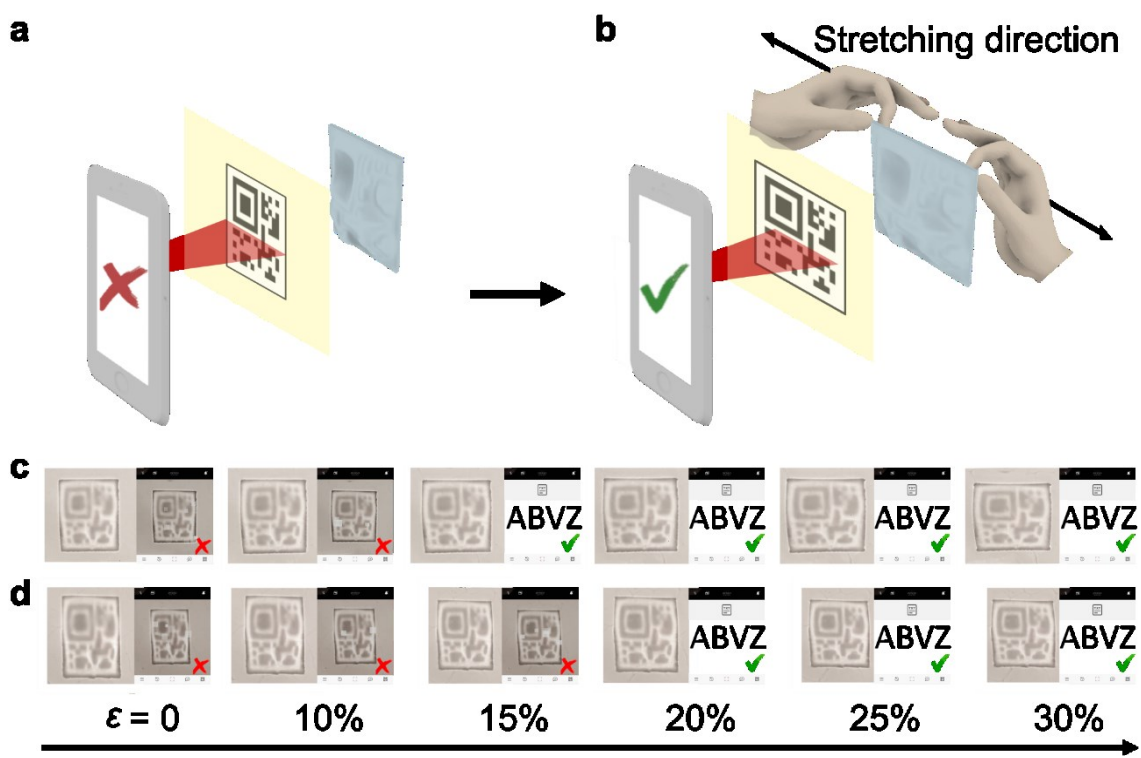

**Figure 7.** Deformable elastomeric MWs. **a,b** Illustration of the working principle of stretchable cryptographic MWs. The obtained light pattern is controlled by the mechanical deformation and made readable by a smartphone scanner only upon applying a calibrated stretching. **c-d** Photographs (left side of the pictures) of the patterns projected by the controllably elongated elastomeric MWs undergoing uniaxial strains, $\varepsilon$, increasing from 0 to 30% from left to right. On the right side of each picture the screenshot of the QR scanner is shown, highlighting images which are decoded by the scanner.

The above results highlight the potential of 3D printing for the versatile manufacturing of cryptographic elements. MWs with surface reliefs are capable of redirecting light rays from sources with different wavelengths, to form pre-determined patterns at a well-defined focal distance. Elastomeric MWs constitute a novel platform combining (i) freeform surface design, (ii) shape and architecture tailorable by controlled mechanical deformation and (iii) optical multi-functionality, encoded through the different 3D shapes that the optical component can acquire in response to an external stimulus (here, uniaxial strain). Such combination results in the possibility to vary the cast patterns, switching from unreadable to readable pictures in a reversible way upon encoding information in the 3D shape of the freeform optical component.

### 3. Conclusion

In summary, we introduced cryptographic, 3D optical components capable of projecting a pre-defined light pattern, upon illumination with light in the visible range, including white sources. Enhanced by inverse Laplacian design and replication methods as done in this work, DLP can enable the fabrication of freeform elements also with stretchable materials.[3,44] We anticipate that 3D printed MWs might be also used for large-scale, broadband beam shaping and control, which is especially useful in those spectral intervals where beam shaping is more demanding, such as for the THz or Mid-IR wavelengths. For such spectral ranges, the larger wavelengths would also lead to less stringent requirements of printing in order to achieve high-quality optical surfaces, and various stereolithographic methods are currently available to this aim.[28] Printable resins with low attenuation losses in these spectral ranges have been recently demonstrated,[45] and could be used for the additive manufacturing of high-quality, freeform optical systems. Stretchable MWs with tailored functionality might find application in fields as various as illumination design, smart labels, anti-counterfeiting, and cryptographic systems.

### 4. Experimental Section

*Calculation of the surface relief of the MW*. Given a generic freeform interface with sufficiently gentle height variation, the intensity profile, $I(R)$, that is generated by the refraction of incident light rays satisfies the relation:[37]

$$I(R) = \frac{1}{1 + f_{MW}(n-1)\nabla^2 h(R)} \tag{4}$$

Equation (4) provides us with a direct relation between the intensity profile of the image to be projected and the height of the surface of the MW, $h(R)$, as reported in Equation (1).

In order to find $h(R)$ one has to solve the inverse Laplacian of Equation (1), which can be done numerically with a finite difference approach:

$$h(i,j) = \frac{1}{4}\left[-\nabla^2 h(i,j)\delta x \delta y + h(i-1,j) + h(i,j-1) + h(i+1,j) + h(i,j+1)\right] \tag{5}$$

where $\delta x = x(i+1) - x(i)$ and $\delta y = y(j+1) - y(j)$, respectively. We point out that the target image is a digital image composed by a finite number of pixels, and $i, j$ are the indexes of the pixel in the $i$-th row and $j$-th column, respectively. The values of the height profile are found starting from the calculation of the right side of Equation (4) at a point $R(i,j) = \{x(i), y(j)\}$. Equation (5) must be solved for every $x(i), y(j)$ and the process must be iterated until it reaches convergence. The boundary conditions are chosen to zero the height of the surface at the edges. Moreover, the intensity of the image must be normalized with the windows area:

$$\int_{\{x,y\}=0}^{\{x,y\}=L} I(x,y) dx dy = A = L^2 \qquad (6)$$

with $L$ the side length of the MW (here assumed to be squared). If this step is not performed the height profile is superimposed on a concave or convex parabolic profile.[37] Indeed, the possibility of modifying the surface relief of the MW, by considering a normalization constant in Eq. (6) $A > L^2$ ($A < L^2$) for a convex (concave) shape gives an additional flexibility for the design of the MW, which can be better tailored to the 3D printing process. In fact, this normalization method is exploited to further engineer the MW and create a convex profile, which is important because the convex shape prevents the air from getting stuck under the previously printed layers, creating bubbles and ripples in the final 3D printed MW (Figure S14). The variation of the normalization constant, $A$, has two consequences: (i) the focal distance will slightly decrease (increase) due to the superimposed convex (concave) profile which changes the angle of incidence on the surface features and (ii) the contrast between black and white areas of the projected intensity profile will slightly decrease as the black regions in the digital image will now become gray.

In order to design the stretchable MW, one has to consider that the replica molding results in a PDMS MW with an inverted surface profile with respect to the 3D printed template. Therefore, the PDMS MW will project the same image but with black and white areas inverted. To realize a PDMS MW that projects the original image, the template MW was designed and 3D printed starting from a negative target image. The flexible MW can be also designed to project a distorted image, for example shrunken in one direction, so that the undistorted, desired image can be obtained by uniaxial stretching of the MW. To this aim, the thinning of the material when stretched has to be accounted for. The relation between the stretched length and the height variation can be derived through the Poisson's ratio, $v$, of the deformable material:

$$v = -\frac{(h-h_0)/h_0}{(L-L_0)/L_0} \tag{7}$$

where $h_0$ and $L_0$ are the thickness and length at rest, $h$ and $L$ are corresponding stretched values, respectively ($v$=0.49 for PDMS).[46] The surface profile of the MW is firstly calculated for the undistorted image, and then shrunk along one axis, e.g. the $x$ axis, by a certain percentage. The thickness of the deformed MW is then derived through Equation (7).

*3D printed and elastomeric MWs*. The MWs are printed by using the E-shell® 600 (ENVISIONTEC®) photo-polymer. The MWs are printed by using the Micro Plus HD (ENVISIONTEC®) system, a DLP 3D printer with a maximum printing volume of 45×28×100 mm$^3$, by setting a thickness of single layers of 15 μm (see Figure S8 and S9). After printing, the MWs are rinsed in isopropanol, dried under a nitrogen flow, and inspected by optical microscopy by using an upright stereomicroscope (MZ16 FA, Leica) and a bright field microscope (BX52, Olympus). In addition, the features of the 3D printed MWs are imaged by scanning electron microscopy (SEM, Merlin, Zeiss) and by an optical profilometer (Talysurf CCI, Taylor Hobson Precision). To this aim, a thin layer of Cr (thickness 20 nm) is thermally evaporated on the MW surface. Optical transmittance spectra are measured by using a Lambda 950 spectrophotometer (Perkin Elmer). Elastomeric MWs are fabricated using a PDMS pre-polymer (Sylgard 184, Dow Corning) by mixing the base and the curing agent with a 9:1 weight:weight ratio. The resulting mixture is degassed in vacuum for 1 hour to remove air bubbles, and poured in a Petri dish containing the 3D printed MWs. Once the crosslinking of the PDMS is completed (typically after 48 hours at room temperature), the replica is peeled off.

*Characterization*. The properties of 3D printed and stretchable MWs are assessed by projection experiments. The printed MWs are mounted on a micrometric translation stage and illuminated with either an incandescent light bulb with variable intensity or a white emitting LED, and with different LEDs with emission wavelength peaked at 475, 524 and 633 nm, respectively. The pattern generated by the MWs is projected on a polyethylene screen (optical transmission spectrum in Figure S15) and captured with a photocamera placed on the opposite side of the screen (Figure S3). To record the intensity distribution of the cast light pattern, the screen is replaced by a CMOS camera (Thorlabs). For testing the MWs under stretching, they are locked between two clamps mounted on micro-translators which allow samples to be stretched along one direction.

**Acknowledgements**
The research leading to these results has received funding from the European Research Council (ERC) under the European Union's Horizon 2020 research and innovation programme (grant agreements No. 682157, "*x*PRINT") and from the Italian Minister of University and Research PRIN 2017PHRM8X project. A. Pitanti and L. Romano are gratefully acknowledged for profilometer characterization and SEM images, respectively.
F. D'Elia and F. Pisani contributed equally to this work.

References

[1] J. Tang, J. Ren, K. Y. Han, *Nanophotonics* **2019**, *8*, 2111.

[2] F. Duerr, Y. Meuret, H. Thienpont, *Opt. Express* **2013**, *21*, 31072.

[3] P. Kunwar, A. V. S. Jannini, Z. Xiong, M. J. Ransbottom, J. S. Perkins, J. H. Henderson, J. M. Hasenwinkel, P. Soman *ACS Appl. Mater. Interfaces* **2020**, *12*, 1640.

[4] F. Nan, Z. Yan, *Nano Lett.* **2019**, *19*, 3353.

[5] Z. Cao, X. Zhang, G. Osnabrugge, J. Li, I. M. Vellekoop, A. M. Koonen, *Light Sci. Appl.* **2019**, *8*, 1.

[6] M. W. Matthès, P. del Hougne, J. de Rosny, G. Lerosey, S. M. Popoff, *Optica* **2019**, *6*, 465.

[7] P. Zupancic, P. M. Preiss, R. Ma, A. Lukin, M. E. Tai, M. Rispoli, R. Islam, M. Greiner, *Opt. Express* **2016**, *24*, 13881.

[8] A. Forbes, A. Dudley, M. McLaren, *Adv. Opt. Photonics* **2016**, *8*, 200.

[9] J. Scheuer, *Nanophotonics* **2017**, *6*, 137.

[10] H. Ries, J. Muschaweck, *J. Opt. Soc. Am. A* **2002**, *19*, 590.

[11] R. Wu, Z. Feng, Z. Zheng, R. Liang, P. Benítez, J. C. Miñano, F. Duerr, *Laser Photonics Rev.* **2018**, *12*, 1700310.

[12] I. Kaya, J. P. Rolland, *Adv. Opt. Technol.* **2013**, *2*, 81.

[13] R. Völkel, M. Eisner, K. J. Weible, *Microelectron. Eng.* **2003**, *67*, 461.

[14] E. Muslimov, E. Hugot, W. Jahn, S. Vives, M. Ferrari, B. Chambion, D. Henry, C. Gaschet, *Opt. Express* **2017**, *25*, 14598.

[15] T. Blalock, K. Medicus, J. D. Nelson, *Proc. SPIE* **2015**, *9575*, 95750H.

[16] R. Wu, L. Xu, P. Liu, Y. Zhang, Z. Zheng, H. Li, X. & Liu, *Opt. Lett.* **2013**, *38*, 229.

[17] D. L. Bourell, *Annu. Rev. Mater. Res.* **2016**, *46*, 1.

[18] J. del Barrio, C. Sánchez-Somolinos, *Adv. Opt. Mater.* **2019**, *7*, 1900598.

[19] A. Camposeo, L. Persano, M. Farsari, D. Pisignano, *Adv. Opt. Mater.* **2019**, *7*, 1800419.

[20] S. C. Ligon, R. Liska, J. Stampfl, M. Gurr, R. Mülhaupt, *Chem. Rev.* **2017**, *117*, 10212.

[21] Z. C. Eckel, C. Zhou, J. H. Martin, A. J. Jacobsen, W. B. Carter, T. A. Schaedler, *Science* **2016**, *351*, 58.

[22] R. L. Truby, J. A. Lewis, *Nature* **2016**, *540*, 371.

[23] H. Y. Jeong, E. Lee, S. An, Y. Lim, Y. C. Jun, *Nanophotonics* **2020**, *9*, 1139.

[24] S. Schmidt, S. Thiele, A. Toulouse, C. Bösel, T. Tiess, A. Herkommer, H. Gross, H. Giessen, *Optica* **2020**, *7*, 1279.

[25] N. Vaidya, O. Solgaard, *Microsyst. Nanoeng.* **2018**, *4*, 18.

[26] J. R. Tumbleston, D. Shirvanyants, N. Ermoshkin, R. Janusziewicz, A. R. Johnson, D. Kelly, K. Chen, R. Pinschmidt, J. P. Rolland, A. Ermoshkin, E. T. Samulski, J. M. DeSimone, *Science* **2015**, *347*, 1349.

[27] X. Chen, W. Liu, B. Dong, J. Lee, H. O. T. Ware, H. F. Zhang, C. Sun, *Adv. Mater.* **2018**, *30*, 1705683.

[28] V. Hahn, P. Kiefer, T. Frenzel, J. Qu, E. Blasco, C. Barner-Kowollik, M. Wegener, *Adv. Function. Mater.* **2020**, *30*, 1907795.

[29] R. Chen, D. Feng, G. Chen, X. Chen, W. Hong, *Adv. Funct. Mater.* **2021**, *31*, 2009916.

[30] W. Ren, G. Lin, C. Clarke, J. Zhou, D. Jin, *Adv. Mater.* **2020**, *32*, 1901430

[31] X. Zheng, Y. Zhu, Y. Liu, L. Zhou, Z. Xu, C. Feng, C. Zheng, Y. Zheng, J. Bai, K. Yang, D. Zhu, J. Yao, H. Hu, Y. Zheng, T. Guo, F. Li, *ACS Appl. Mater. Interfaces 2021*, **13**, 15701.

[32] H.-Y. Wang, K.-K. Yu, C.-Y. Tan, K. Li, Y.-H. Liu, L. Shi, K. Lu, X.-Q. Yu, *J. Mater. Chem. C* **2021**, *9*, 2864

[33] S. Geiger, J. Michon, S. Liu, J. Qin, J. Ni, J. Hu, T. Gu, N. Lu, *ACS Photonics* **2020**, *7*, 2618.

[34] S. C. Malek, H. S. Ee, R. Agarwal, *Nano Lett.* **2017**, *17*, 3641.

[35] J. Wang, Y. Yin, Q. Hao, S. Huang, E. Saei Ghareh Naz, O. G. Schmidt, L. Ma, *ACS Photonics* **2018**, *5*, 2060.

[36] S. Petsch, S. Schuhladen, L. Dreesen, H. Zappe, *Light: Sci. Appl.* **2016**, *5*, e16068.

[37] M. V. Berry, *J. Opt*. **2017**, *19*, 06LT01.

[38] F. Aloui, L. Lecamp, P. Lebaudy, F. Burel, *Express Pol. Lett.* **2018**, *12*, 966.

[39] W. B. Thompson, G. E. Legge, D. J. Kersten, R. A. Shakespeare, Q. Lei, *J. Opt. Soc. Am. A* **2017**, *34*, 583.

[40] R. N. Bryan, *Introduction to the science of medical imaging*, Cambridge Univ. Press, **2009**.

[41] J. Fischer, M. Wegener, *Laser Photonics Rev.* **2013**, *7*, 22.

[42] L. Jeng-An, F. Chiou-Shann, *Math. Probl. Eng.* **2013**, *2013*, 1.

[43] L. Karrach, E. Pivarčiová, P. Božek, *J. Imaging* **2020**, *6*, 67.

[44] Z. Hong, R. Liang, *Sci. Rep.* **2017**, *7*, 1.

[45] F. Zhou, W. Cao, B. Dong, T. Reissman, W. Zhang, C. Sun, *Adv. Opt. Mater.* **2016**, *4*, 1034.

[46] A. Müller, M. C. Wapler, U. Wallrabe, *Soft Matter* **2019**, *15*, 779.

# Supporting Information

**Cryptographic Strain-Dependent Light Pattern Generators**

*Francesca D'Elia,[1] Francesco Pisani,[2]\* Alessandro Tredicucci,[2,3] Dario Pisignano,[2,3] Andrea Camposeo[3]\**

[1]NEST, Scuola Normale Superiore, Piazza S. Silvestro 12, I-56127 Pisa, Italy

[2]Dipartimento di Fisica, Università di Pisa, Largo B. Pontecorvo 3, I-56127 Pisa, Italy

E-mail: francesco.pisani@df.unipi.it

[3]NEST, Istituto Nanoscienze-CNR and Scuola Normale Superiore, Piazza S. Silvestro 12,

I-56127 Pisa, Italy

E-mail: andrea.camposeo@nano.cnr.it

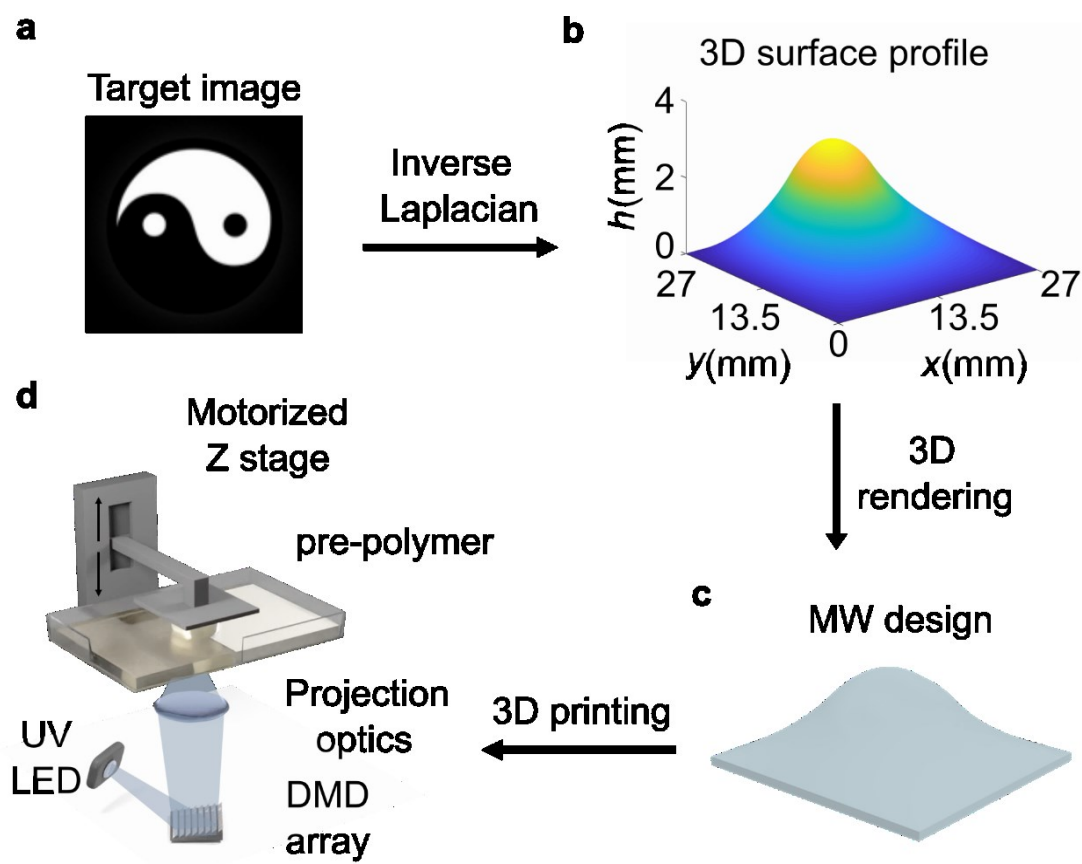

**Figure S1**. Schematic illustration of the design of the MWs. The starting point is the target pattern to be projected by the MW (exemplary target pattern shown in **a**). **b** Corresponding 3D surface profile of the MW, calculated by numerically solving the inverse Laplacian of the target intensity profile. **c** CAD model of the MW, generated by rendering of the surface profile shown in **b**. A base plate of thickness 1 mm is added to the surface profile shown in **b**. **d** Realization by 3D printing with a digital light processing system.

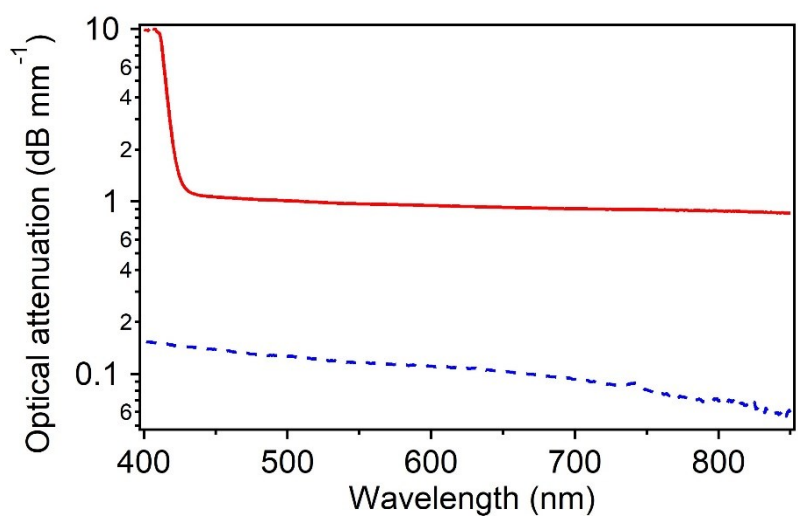

**Figure S2**. Spectrum of the optical attenuation for 3D printed E-Shell® 600 (red continuous line) and PDMS (blue dashed line). The data are obtained by measuring the transmittance spectrum of samples with thickness of 4 mm. The data include Fresnel losses.

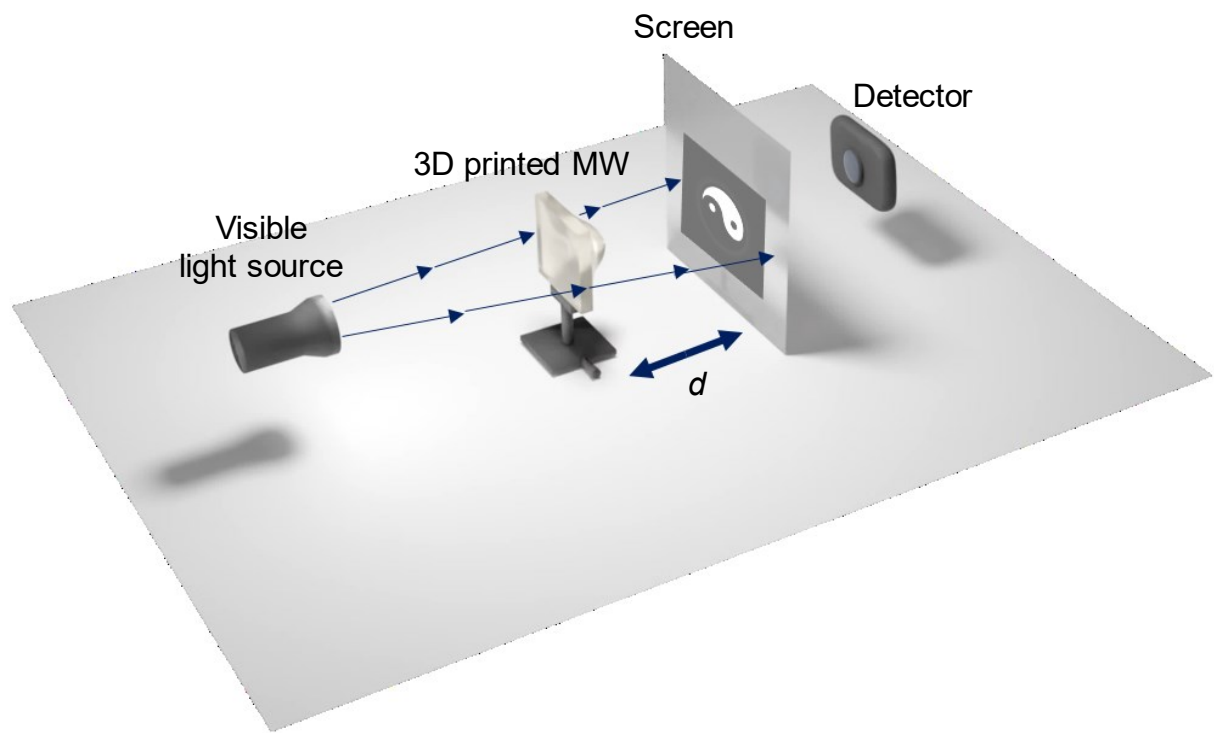

**Figure S3**. Schematic illustration of the experimental set-up used for the characterization of the light patterns generated by the magic windows (MWs). The MWs are illuminated by various visible light source (an incandescent lamp with variable intensity, a white LED, and several LEDs with emission peaked at 478 nm, 555 nm, and 633 nm, respectively). The samples are mounted on a micrometric translation stage. The rays refracted by the MWs are collected on a screen placed at a distance, *d*, and captured with a photocamera placed on the opposite side of the screen. Alternatively, the screen can be removed and the intensity of the light pattern generated by the MW can be measured by using a CMOS camera detector.

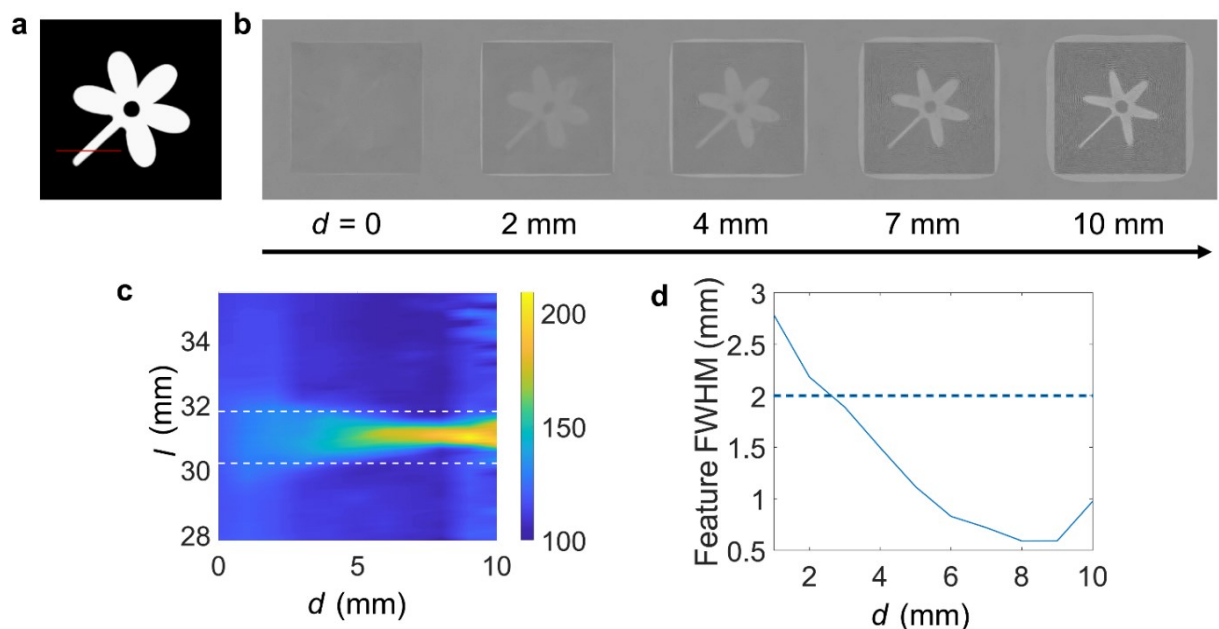

**Figure S4**. **a** Target image. **b** Simulated intensity patterns generated by the MW at various distances, *d*. The intensity maps are obtained by ray tracing simulations performed by Blender Luxcore. The 3D MW is illuminated with a plane wave and the light refracted is collected on a semi-transparent screen, similarly to the experimental set-up used for the characterization of the printed optical components (Figure S2). The refractive index of the MW is 1.5, and the MW is transparent (diffusion or scattering of light inside the MW are not accounted for) and smooth in the simulation. **c** Intensity profile *vs.* the coordinate, *l*, along the red line shown in the target image (**a**) at various distances, *d*. The dashed lines correspond to the stem size of the target image. **d** Stem size (obtained as the FWHM of the light intensity profiles shown in **c**) *vs.* *d*. The horizontal dashed line marks the size of the flower stem of the target image. Optical imaging, corresponding to the focal distance, is obtained with *d* between 2 and 3 mm.

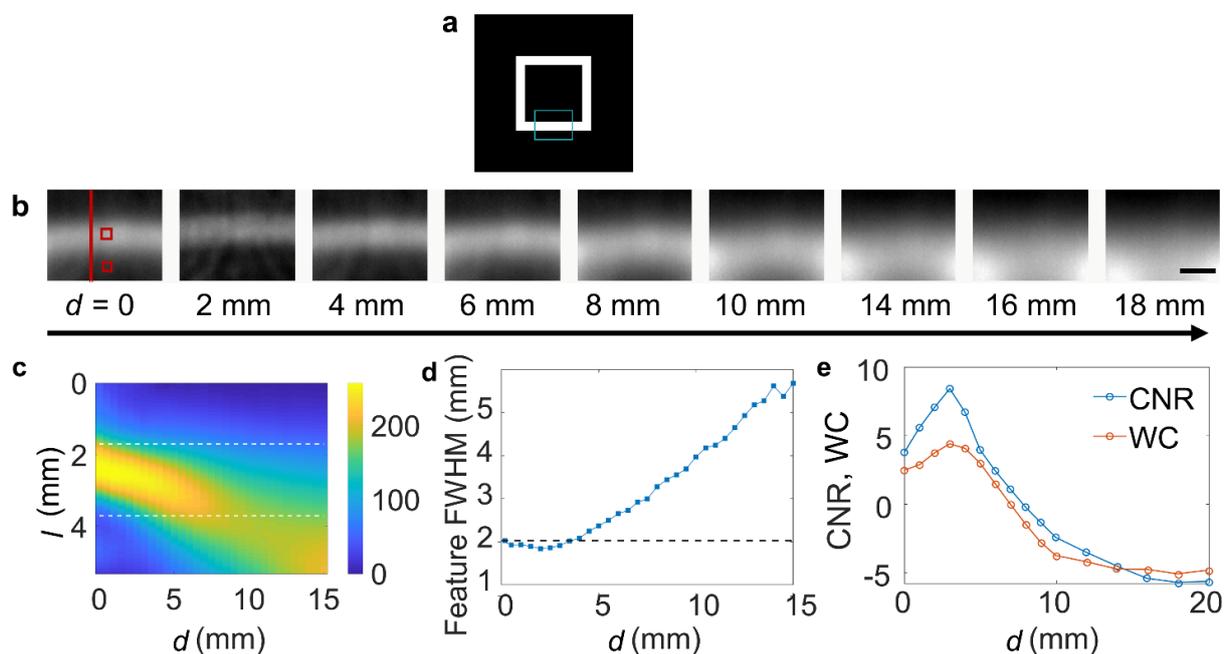

**Figure S5**. **Quantitative analysis of the intensity patterns generated by a MW. a** Target image. The feature investigated is highlighted by a blue square. **b** Maps of the corresponding, experimental intensity acquired by positioning a CMOS camera at different distances, *d*. Scale bar: 2 mm. **c** 2-dimensional plot of the light intensity profile along the red line in (**b**). The dashed lines highlight the expected width of the white area in (**a**). **d** Variation of the width of the white area (estimated as the FWHM of the intensity profiles in (**c**) vs. *d*. The dashed horizontal line marks the expected width of the white area in (**a**). **e** CNR and WC parameters vs. *d*. CNR and WC are calculated using the areas highlighted with red boxes in (**b**).

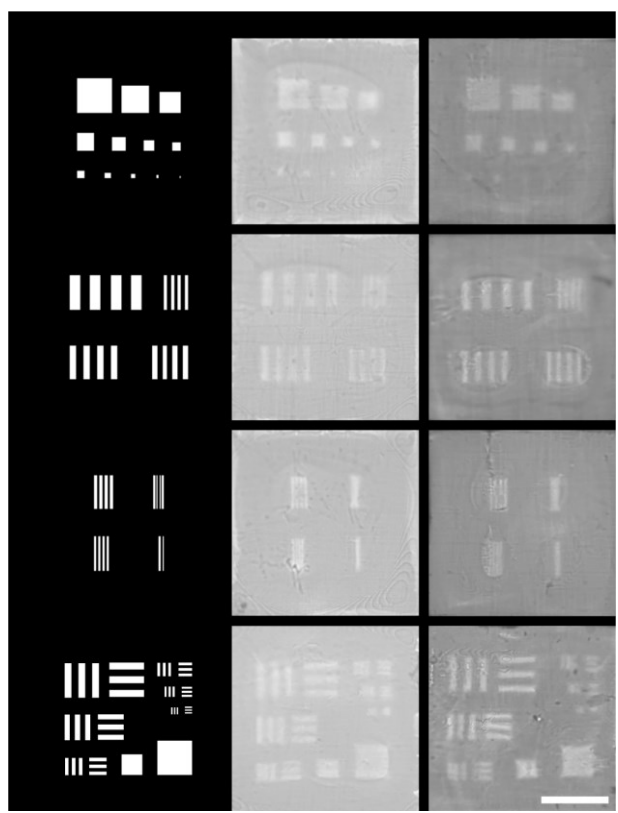

**Figure S6**. MW resolution tests. **a-d** Target images (left side) used for evaluating the minimum feature size that can be projected by 3D printed MWs (central image) and by elastomeric MWs (right side). The projected images are obtained upon illumination with a W lamp. Scale bar: 1 cm. Features shown in **a** are squares with side ranging from 5 mm (top-left square) down to 200 µm (bottom-right square). Features shown in **b** are lines with spacing and feature size of 1.5, 1, 1.25 and 0.75 mm, whereas features shown in **c** are lines with spacing and feature size of 400, 300, 200, 100 µm. The patterns shown in **d** are squares with side of 5 and 3 mm, respectively, and groups of vertical and horizontal lines with linewidth: 1, 0.75, 0.5, 0.4, 0.3, 0.2 mm.

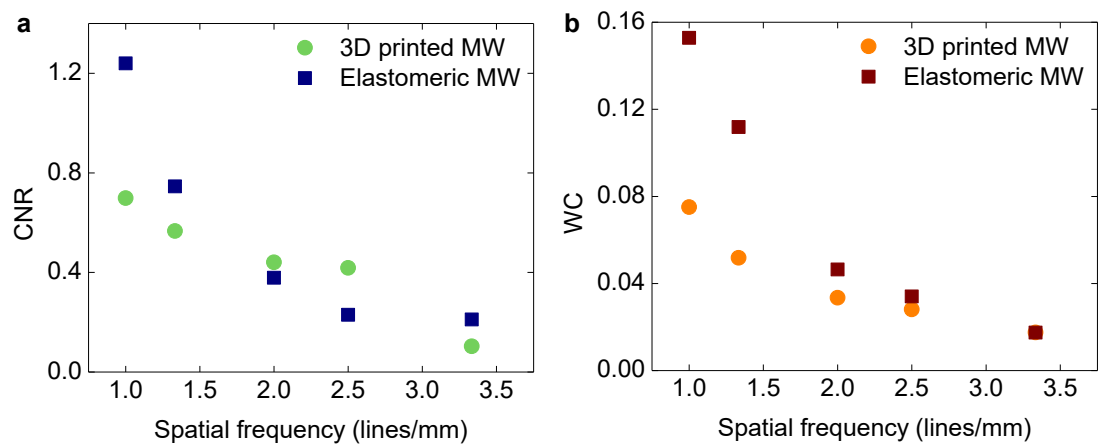

**Figure S7**. Contrast analysis performed by calculating the CNR (**a**) and the Weber, WC, (**b**) parameters (Equations (2) and (3) of the main text, respectively) utilizing the images projected by the 3D printed (circles) and elastomeric (squares) MWs (Figure S6d).

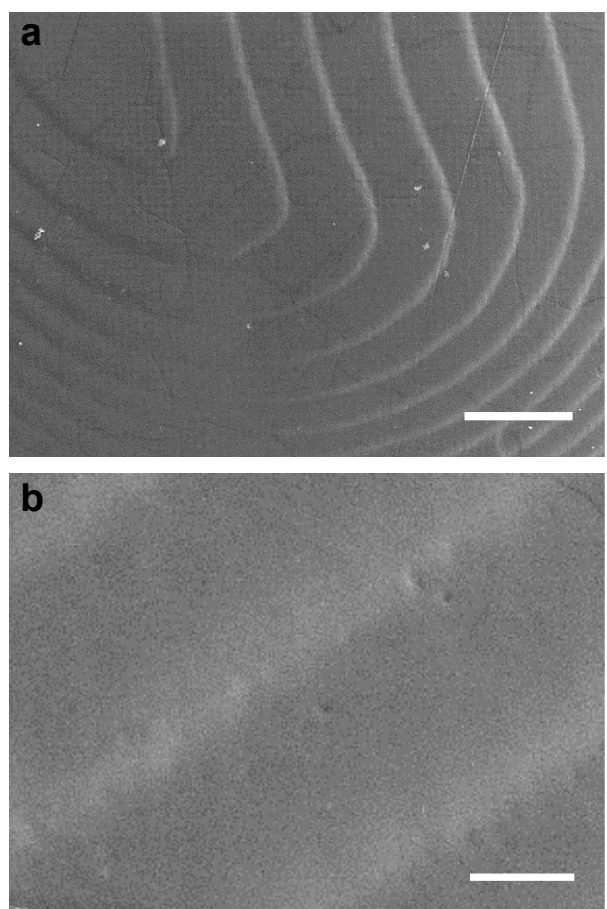

**Figure S8**. Scanning electron microscopy (SEM) on 3D printed MW. **a-b** SEM micrographs of the surface of the MW generating the flower pattern, at different magnifications. Scale bars: 500 μm (**a**) and 50 μm (**b**).

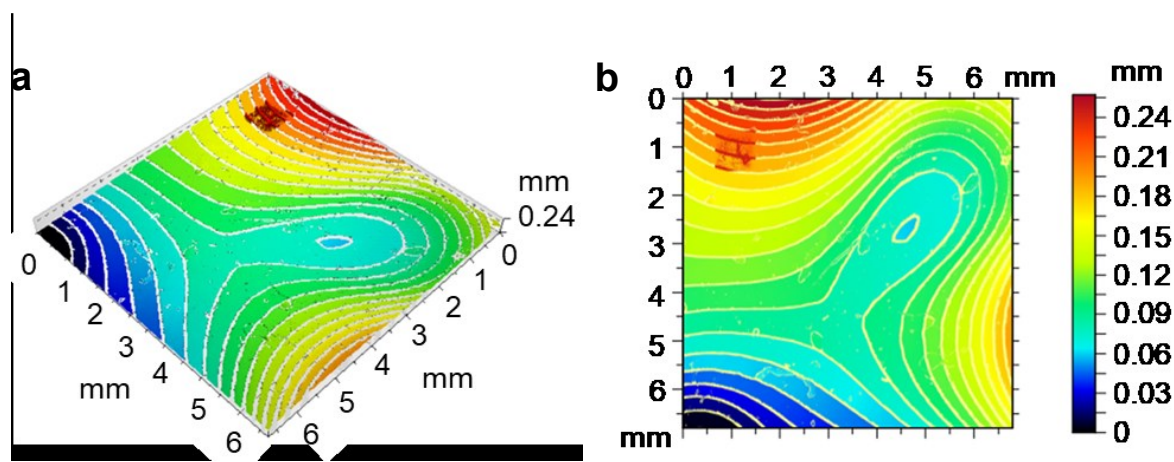

**Figure S9**. **a-b** 3-dimensional (**a**) and 2-dimensional (**b**) maps obtained by means of an optical profilometer of the surface of a MW, highlighting the various printed layers with thickness 15 μm.

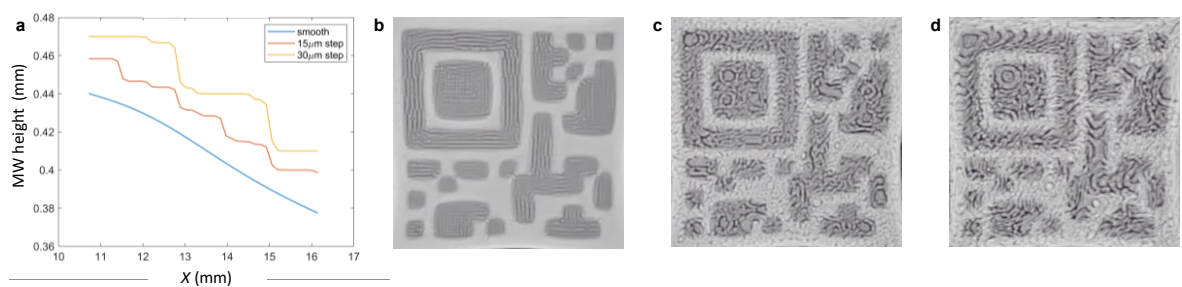

**Figure S10**. **a** Example of the height profile of part of a MW for a smooth surface (blue line) and for stepped surface with step height of 15 μm (orange line) and 30 μm (yellow line). The height profiles are vertical shifted for better clarity. **b-d** Simulated intensity patterns generated by MW with a smooth surface (**b**) and with a stepped surface with step height of 15 μm (**c**) and 30 μm (**d**).

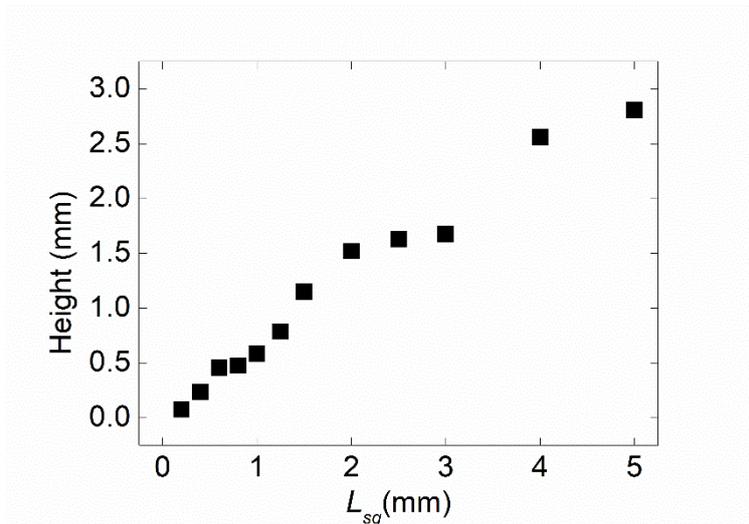

**Figure S11**. Dependence of the height of the surface profile of the squares shown in Figure S6a on the side of the square ($L_{sq}$).

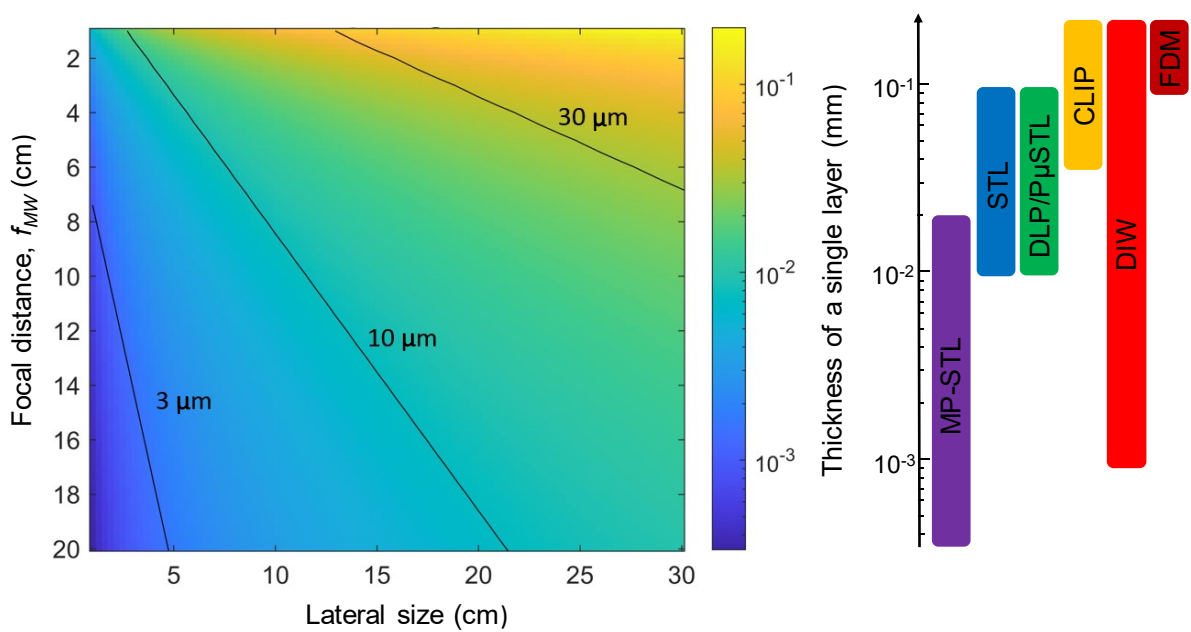

**Figure S12**. The color map on the left shows the dependence of the single layer thickness for a MW that is sliced in 100 layers, as a function of the focal distance and the lateral size of the MW. The thickness of each layer is proportional to the lateral size of the MW and inversely proportional to the focal distance. The continuous lines mark the intervals with a thickness of the layers of 3, 10 and 30 µm. Right side: interval of spatial resolution of 3D printing technologies that match the desired single-layer thickness (vertical scale). The data are from References [S1-S3]. MP-STL: multiphoton stereolitography, STL: stereolitography; DLP: digital light processing; PµSTL: projection micro-stereolitography; CLIP: continuous liquid interface printing; DIW: direct ink writing; FDM: fused deposition modeling.

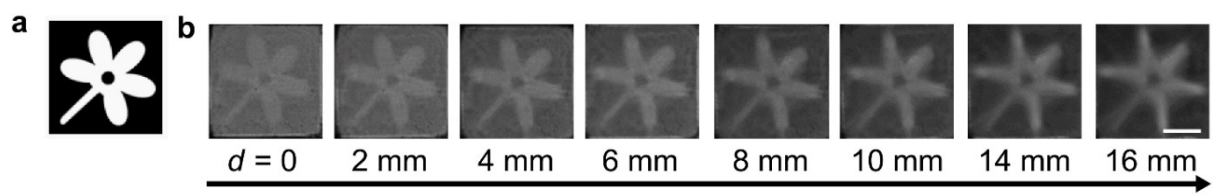

**Figure S13**. **a** Target image used to design and realize the MW. **b** Photographs of the light patterns projected on the screen positioned at different distances, *d*, from the MW made of PDMS. Scale bar: 1 cm.

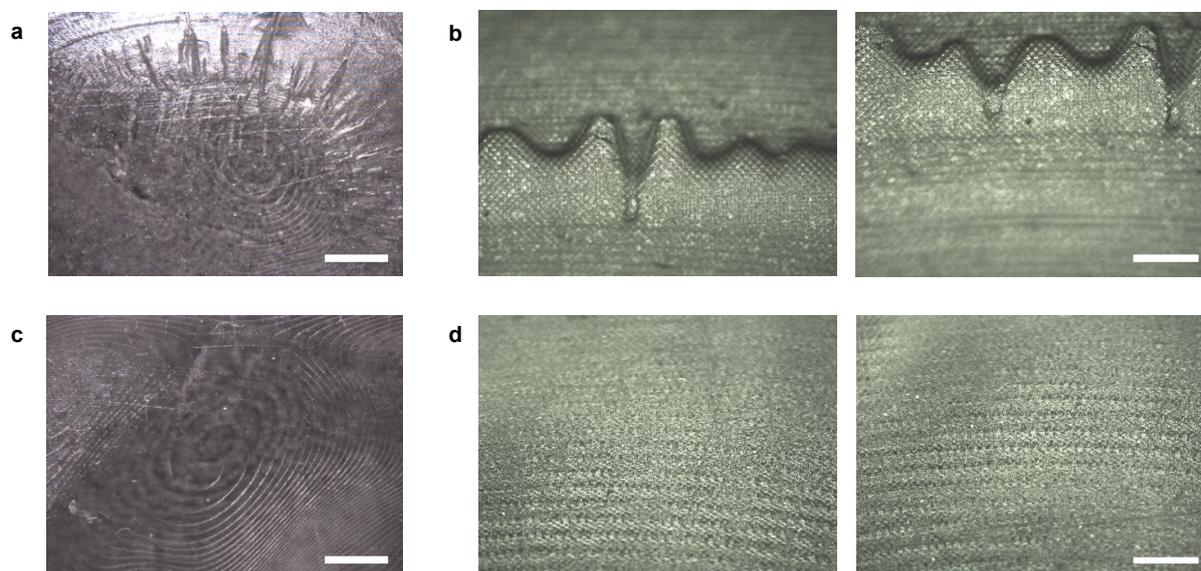

**Figure S14**. **a-d** Optical micrographs of the MW generating the Yin Yang symbol with a concave (**a**,**b**) and convex (**c**,**d**) surface profile. Scale bars: 2 mm (**a**,**c**) and 400 μm (**b**,**d**).

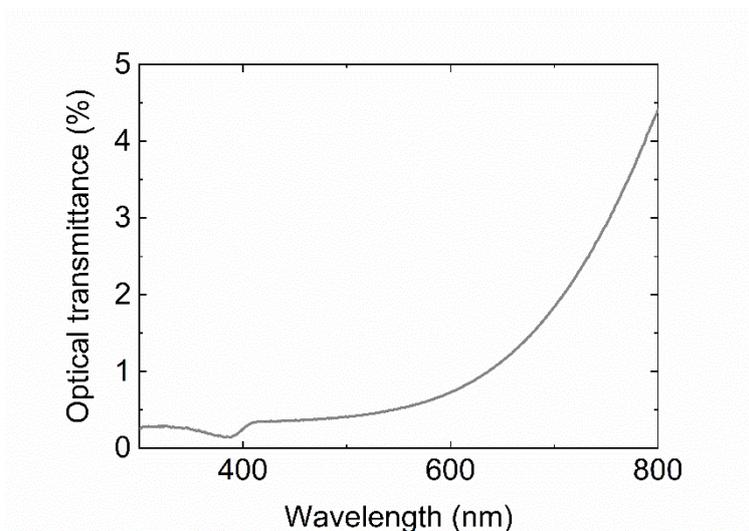

**Figure S15**. Spectrum of the MW-generated light intensity, transmitted by a polyethylene screen used for imaging.

**References**

[S1] M. Shusteff, A. E. M. Browar, B. E. Kelly, J. Henriksson, T. H. Weisgraber, R. M. Panas, N. X. Fang, C. M. Spadaccini, *Sci. Adv.* **2017**, *3*, eaao5496.

[S2] A. Camposeo, L. Persano, M. Farsari, D. Pisignano, *Adv. Opt. Mater*. **2019**, *7*, 1800419.

[S3] V. Hahn, P. Kiefer, T. Frenzel, J. Qu, E. Blasco, C. Barner-Kowollik, M. Wegener, Adv. Funct. Mater. 2020, 30, 1907795.